\documentclass[aps,showpacs,showkeys,nofootinbib,preprint]{revtex4}
\usepackage{epsfig,amssymb,bm,graphicx,color,amsmath,rotating}
\usepackage[percent]{overpic}
\usepackage{mwe,xcolor}
\usepackage{transparent}
\newcommand{\nlC}{n$\ell$C\,}
\newcommand{\nlBW}{n$\ell$BW\,}
\begin{document} {\normalsize }

\title{Non-perturbative signatures of non-linear Compton scattering}
%

\author{U.~Hernandez Acosta, A.~Otto, B.~K\"ampfer}
 \affiliation{
 Helmholtz-Zentrum  Dresden-Rossendorf, 01314 Dresden, Germany}
 \affiliation{
 Institut f\"ur Theoretische Physik, TU~Dresden, 01062 Dresden, Germany}

 \author{A.~I.~Titov}
 \affiliation{
 Bogoliubov Laboratory of Theoretical Physics, JINR, Dubna 141980, Russia}

\begin{abstract}
The probabilities of various elementary laser - photon - electron/positron
interactions display in selected phase space and parameter regions
typical non-perturbative dependencies such as 
$\propto {\cal P} \exp\{- a E_{crit} /E\}$, where
${\cal P}$  is a pre-exponential factor, $E_{crit}$ denotes the critical Sauter-Schwinger
field strength, and $E$ characterizes the (laser) field strength.
While the Schwinger process with $a = a_S \equiv \pi$ and the non-linear
Breit-Wheeler process in the tunneling regime with 
$a = a_{n \ell BW} \equiv 4 m / 3 \omega'$ 
(with $\omega'$ the probe photon energy and $m$ the electron/positron mass) 
are famous results, we point out
here that also the non-linear Compton scattering exhibits a similar behavior
when focusing on high harmonics. 
Using a suitable cut-off $c > 0$, the factor $a$ becomes
$a = a_{n \ell C} \equiv \frac23 c m /(p_0 + \sqrt{p_0^2 -m^2)}$. 
This opens the avenue towards a new
signature of the boiling point of the vacuum even for field strengths $E$
below $E_{crit}$ by employing a high electron beam-energy $p_0$
to counter balance the large ratio $E_{crit} / E$ by a small factor $a$
to achieve $E / a \to E_{crit}$.
In the weak-field regime, the cut-off facilitates a threshold leading to
multi-photon signatures showing up in the total cross section at
sub-threshold energies.  
\end{abstract}


\date{\today}

\maketitle

\section{Introduction}

The Schwinger process signals the instability of the vacuum against particle
(pair) creation in an external field. The pair ($e^+ e^-$) production rate
$\propto \exp\{- a E_{crit} / E \}$, $a = \pi$ \cite{Sauter:1931zz,Heisenberg:1935qt,Schwinger:1951nm},
in a spatio-temporally homogeneous electric field of strength $E$
is exceedingly small due to the large value of
the (critical) Sauter-Schwinger field strength $E_{crit} = 1.3 \times 10^{18}$ V/m
and therefore escaped a direct experimental verification until now.
Much hope was therefore put on the progressing laser technology which
however delivers even at present and near-future "ultrahigh intensities" 
far too low field strengths \cite{DiPiazza:2011tq,Tajima}.
Many efforts on the theory side attempted to find field configurations which
enhance the Schwinger type pair production. 
To cite a few entries of the fairly extended literature, which documents
the ongoing enormous interest in that topic,
we mention dynamical assistance \cite{Torgrimsson:2017cyb,Torgrimsson:2017pzs,Fey:2011if,Dunne:2009gi,
Schutzhold:2008pz,Aleksandrov:2016lxd,Aleksandrov:2018uqb,Orthaber:2011cm,
Otto:2016xpn,Panferov:2015yda,Otto:2015gla,Otto:2014ssa,Otto:2016fdo},
double assistance effects \cite{Torgrimsson:2016ant,Otto:2018jbs}, 
multi-beam configurations \cite{Bulanov:2010ei} and their embedding into
optimization procedures \cite{Kohlfurst:2012rb,Hebenstreit:2014lra}. 
In essence, these attempts envisage a reduction
of the factor $a$ in the above exponent, which is in general a complicated 
function of the external parameters. Despite such a "practical goal", these
investigations aim at understanding the QED as a pillar of the Standard Model
in the non-perturbative, high-intensity regime. Given the seminal meaning
of the Schwinger process as paradigm for related processes, e.g.\ particle
production in cosmology \cite{Birrell_Davies} and at black hole horizons as 
Hawking radiation \cite{Hawking:1974sw},  
up to the disputed Unruh radiation
\cite{Unruh:1976db,Crispino:2007eb,Narozhny:2004xx}, 
various authors considered
analog processes, e.g.\ in condensed matter physics 
\cite{Linder:2015fba,Dreisow:2012sx} 
and in wave guides \cite{Lang:2018tpj} etc., 
which display also the monomonial, genuinely non-perturbative
dependence on an external field parameter.    

Still within QED, one can search for more easily accessible processes which
have the prototypical non-perturbative dependence 
$\propto \exp\{- a E_{crit} / E \}$.  
For instance, the LUXE collaboration
\cite{Hartin:2018sha,Altarelli:2019zea,Abramowicz:2019gvx} 
envisages to exploit the non-linear Breit-Wheeler process which is known
to behave as $\propto \exp\{- a_{n\ell BW} E_{crit} / E \}$
in the tunneling regime with $a_{n\ell BW} = 4 m / 3 \omega'$,
where $\omega'$ is the energy of a probe photon traversing a strong
laser pulse. 
LUXE is planed as next-generation follow-up of the seminal SLAC
experiment E-144 \cite{Burke:1997ew}, which operated in the multi-photon regime, by 
"Measuring the Boiling Point of the Vacuum of Quantum Electrodynamics"
\cite{Hartin:2018sha}
via the non-linear Breit-Wheeler process since $\omega' \gg m$ 
reduces the exponential suppression, i.e.\ it makes the above quantity 
$a_{n\ell BW}$ small
when using probe photon energies $\omega'$ much larger 
than the electron mass $m$, thus compensating the large value
of $E_{crit} /E$ at presently attainable facilities.
Note furthermore that the trident process shows also an exponential
behavior under certain conditions \cite{Dinu:2017uoj,Dinu:2019wdw},
as originally elaborated in \cite{Baier:1972vc,Ritus:1972nf}.

Here, we point out that the non-linear Compton process has a similar
non--perturbative exponential field strength dependence under certain side conditions.
The key is the suppression of the low harmonics which facilitate
the Thomson limit and display a polynomial dependence. What is then
left is the otherwise exponentially suppressed contribution. 
The analogy to the non-linear Breit-Wheeler process is not surprising
since it is the crossing channel of the non-linear Compton process 
in the Furry picture. The crucial difference is in the final-state phase
spaces. This is most clearly evident in the perturbative, weak-field
limit, where the Breit-Wheeler process is a threshold process,
while the Compton process without side conditions has no threshold
(see \cite{LL,Harvey:2009ry} for the physical regions 
in the Mandelstam plane).
We introduce here as side condition a cut-off which is related to exit channel
kinematics. This in fact enforces the exponential behavior. 

Our brief note is organized as follows. In section \ref{sect:2}, we outline the
definition of a Lorentz invariant cut-off in the non-linear Compton scattering.
In section \ref{sect:3}, the restriction of the physically accessible regions in
the Mandelstam plane is discussed.
The cut-off facilitates a clear signature of multi-photon effects in the
total cross section in the weak-field regime (section \ref{sect:4}).
The moderately strong-field regime is considered in section \ref{sect:5}, 
where we compare the exact numerical results with some approximation
formula to evidence the exponential dependence of the cross section.  
The discussion section \ref{sect:6} contains a comparison with 
laser pulses and outlines of how the cut-off is realized by photon
observables in the exit channel.
We summarize in section \ref{sect:7}.

\section{Non-linear Compton scattering with cut-off}\label{sect:2}

We consider here a monochromatic laser field in plane wave approximation
for circular polarization.
The non-linear Compton (\nlC) cross section
with cut-off $c$ reads
\begin{equation} \label{sigma}
\sigma = \frac{ \alpha^2 \pi}{a_0^2} \frac{1}{k \cdot p}
\, F (a_0, k\cdot p, c), 
\quad
F (a_0, k \cdot p, c)  = 
\sum_{n = 1}^\infty \int_{c}^{y_n} dx \, \frac{1}{(1+x)^2} \, F_n(z_n), 
\end{equation}
where 
\begin{equation} \label{F}
F_n (z_n) = - 4 J_n(z_n)^2 + 
\left(2 + \frac{x^2}{1+x} \right) 
a_0^2 [J_{n+1}(z_n)^2 + J_{n-1}(z_n)^2 - 2 J_n(z_n)^2]
\end{equation}
for $c \le y_n$ and $F_n = 0$ elsewhere.
The Lorentz and gauge invariant quantity $a_0$ is 
the classical non-linearity parameter characterizing solely the laser beam,
and $\alpha$ stands for the fine-structure constant.  
The arguments of the Bessel functions $J_n$ read explicitly
$z_n(x, y_n, a_0) = 2 n a_0 \frac{1}{y_n}\sqrt{\frac{x (y_n - x)}{1 + a_0^2}}$,
where the two invariants 
$x = k \cdot k' / k \cdot p'$ and $y_n = 2 n \frac{k \cdot p}{m_*^2}$
with $0 \le x \le y_n$ enter. 
For $c = 0$, one recovers the text book formulas, e.g.\  in \cite{LL,Harvey:2009ry},
where the effective mass $m_*^2 = m^2 (1 + a_0^2)$ and the
(quasi-) momentum balance as well as the relation to asymptotic four-momenta
($p/p'$ for $in/out$-electrons and $k/k'$ for $in/out$-photons) are discussed in detail.
The only but decisive difference is the introduction of the cut-off $c$ 
in (\ref{sigma}) which pushes the
lower limit of the $x$ integration to higher values, i.e.\ it is aimed
at suppressing the lower harmonics.

\section{Kinematics in the Mandelstam plane}\label{sect:3}

\begin{figure}[tb!] 
\includegraphics[width=0.77\columnwidth]{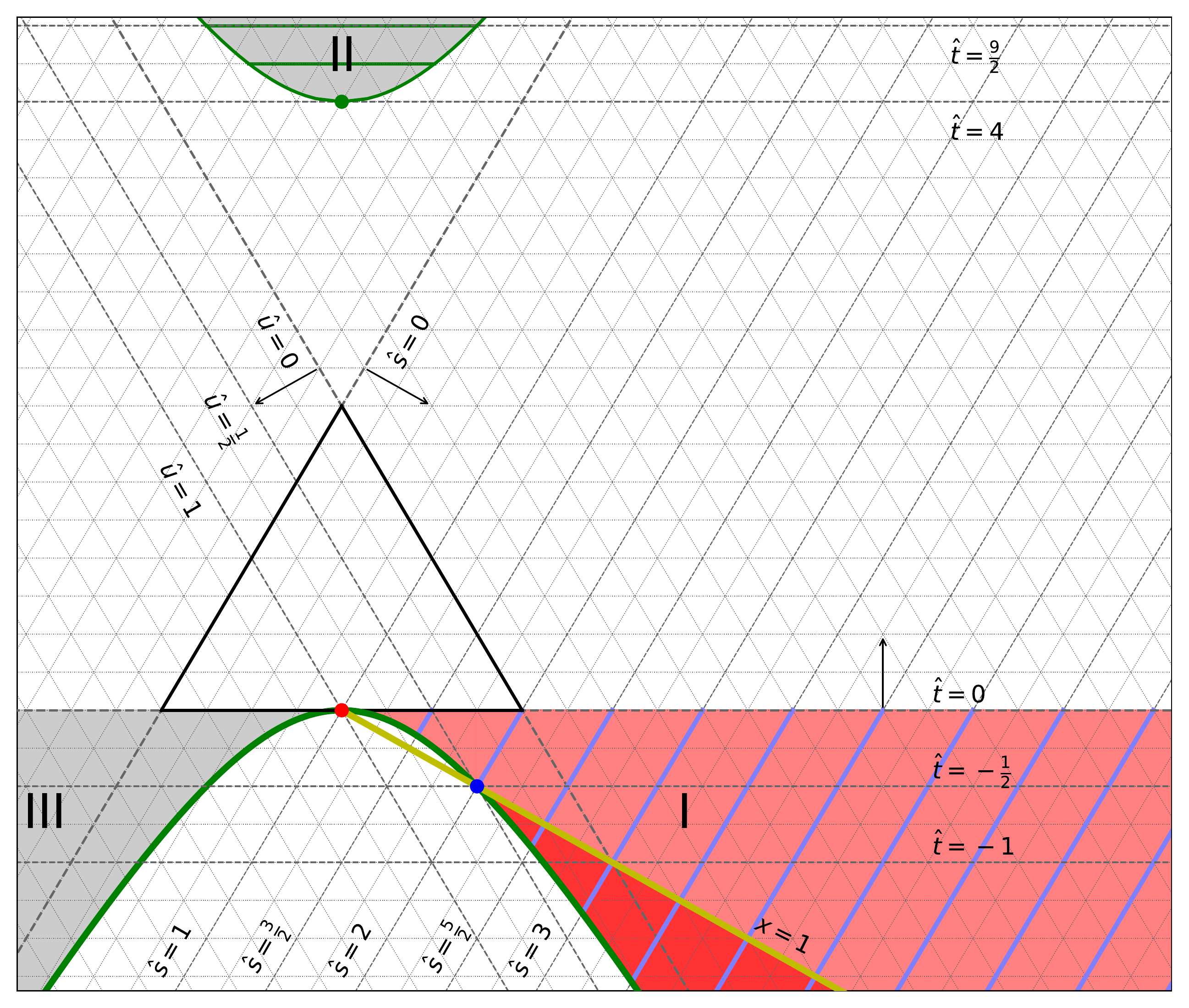}
\put(-182,33){\rotatebox{-40}{{\large $\bm{x > 1}$}}}
\vspace*{-3mm}
\caption{(Color online)
Mandelstam plane with physical regions of 
\nlC process (I, red area), 
\nlBW process (II, upper gray area)
and the mirrored Compton I (III, left gray area).
Arrows indicate the directions of positive variables
$\hat s$, $\hat t$ and $\hat u$, and a few coordinate
values are provided too.
Bullets depict thresholds. Harmonics of \nlC are parallel to (and
may coincide with) the blue lines in region I, which become
restricted to the dark-red region (labeled by ``$x > 1$") 
below the boundary
$x = 1$ (in yellow) and above the hyperbola $\hat s \hat u = 1$ (in green)
when imposing the cut-off $c = 1$ which facilitates the threshold
at coordinates $\hat s = 2$, $\hat t = - \frac12$, $\hat u = \frac12$.
In region II, the harmonics of \nlBW are parallel to (and may coincide with)
the horizontal green lines.
\label{Fig:M}}
\end{figure}

The meaning of the cut-off $c$ can be visualized in a covariant manner by 
inspecting the Mandelstam plane. Defining the invariants
$s_n = (q+ nk)^2$, $t_n = (k' - nk)^2$, $u_n =(q' - nk)^2$ for harmonics
$n = 1, 2, 3 \cdots$, the physical regions I - III in scaled triangular coordinates
$\hat s = s / m_*^2$, $\hat t = t / m_*^2$, $\hat u = u / m_*^2$
with $\hat s + \hat t + \hat u = 2$ refer to processes related by crossing 
symmetry on amplitude level: 
I (red area in Fig.~\ref{Fig:M}) for \nlC
process, $e^- + n \gamma \to {e^-}'  + \gamma'$ or
$q + nk = q' + k'$ with quasi-momenta $q$ and $q'$,
II (upper gray area) for non-linear Breit-Wheeler (\nlBW) pair production,
$\gamma'  + n\gamma \to e^+ + e^-$ or $k' + n k = q_{e^+} + q_{e^-}$, 
and III (left gray area) as mirror of I, e.g.\
$e^+ + n \gamma \to {e^+}' + \gamma'$.
In I, the harmonics $\hat s_n = const$ are parallel lines (in blue in Fig.~\ref{Fig:M}),
limited by $\hat t = 0$ (on-axis forward scattering, where $x =0$) and by the hyperbola
$\hat s \hat u = 1$ (on-axis backscattering), i.e.\ the physical interval 
of each harmonic is given by $0 \le \hat t \le 2 - \hat s_n - \hat s_n^{-1}$,
which is another way of expressing the above quoted restriction $0\le x \le y_n$. 
The scaled invariant-energy squared of the first harmonic is 
$\hat s_1 = 1 + \Delta \hat s $ 
(measured from the bullet at the top of I in direction of the $\hat s$
coordinate, indicated by the arrow, as shown for the other coordinates too)
and the spacing of adjacent harmonics is 
$\Delta \hat s=\hat s_{n+1} - \hat s_n = 2 k \cdot p / m_*^2$.
Considering an optical laser 
(we use the frequency $\omega = 1$~eV as representative value) 
colliding head-on with an electron beam, 
as available (i) in HZDR (40 MeV \cite{Jochmann:2013toa}) 
or planned (ii) at ELI (600 MeV \cite{Turcu:2016dxm}) and
(iii) at LUXE (17.5 GeV \cite{Altarelli:2019zea})
for instance, one has 
(i) $\Delta \hat s \approx 6.5 \times 10^{-3} /(1+a_0^2)$,
(ii)  $9.6 \times 10^{-2} /(1+ a_0^2)$ and 
(iii) $2.8 \times 10^{-1} /(1+ a_0^2)$
in the red region displayed in Fig.~\ref{Fig:M}.
Instead of displaying so many narrow parallel lines
representing the harmonics, we depict only a 
few representative proxies of them at $\hat s = \frac32, 2, \frac52, 3$ etc.\ 
as blue lines.
In contrast to the perturbative, weak-field limits of the linear processes,
$n = 1, a_0 \to 0$, the physical regions I - III of the non-linear processes
are mapped out by the discrete harmonics $n = 1 \cdots \infty$.

The cut-off $c = 1$ in (\ref{sigma}) restricts the region I to the dark-red
area, limited by a section of the hyperbola $\hat s \hat u = 1$ and the line
$x \equiv k \cdot k' / k \cdot p' = \hat t_n /(1 - \hat s_n - \hat t_n) \ge c$. 
This excludes the 
low harmonics $\hat s_n < 2$ and restricts the admissible $\hat t$ intervals
of the harmonics $\hat s_n \ge 2$ to
$\frac12 (1 -\hat s_n) \ge \hat t \ge 2 - \hat s_n - \hat s_n^{-1}$.
For the above quoted numbers, harmonics with 
$n > (1+ a_0^2) / 2 p \cdot k$ are in the admissible region. 
In such a way, a non-trivial threshold
is introduced, depicted by the blue bullet at the tip of the dark-red area at coordinates
$\hat s = 2$, $\hat t = - \frac12$ and $\hat u = \frac12$. 
Imagine now that we keep the laser
frequency $\omega = \vert \vec k \vert$ 
but lower the electron energy $p_0$, i.e.\ the values
of $\hat s_n$ would become gradually smaller. Then, a certain number of harmonics
drop out the admissible area as they pass the threshold by moving to the left-above:
less and less harmonics contribute to the \nlC
process by (i) imposing a threshold by the cut-off $c > 0$ and/or
(ii) diminishing $\hat s_1$ (and all other $\hat s_n$). 

\section{Multi-photon regime, $\mathbf{a_0 < 1}$}\label{sect:4}

\begin{figure}[tb!] 
\includegraphics[width=0.49\columnwidth]{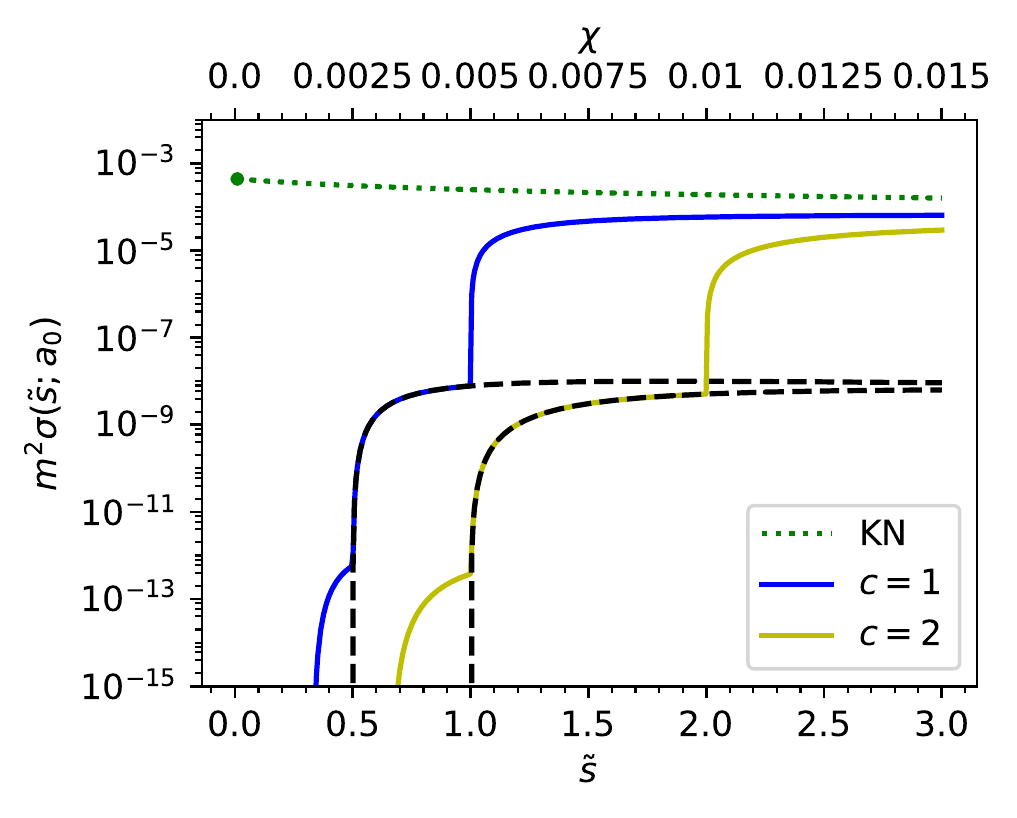}
\includegraphics[width=0.49\columnwidth]{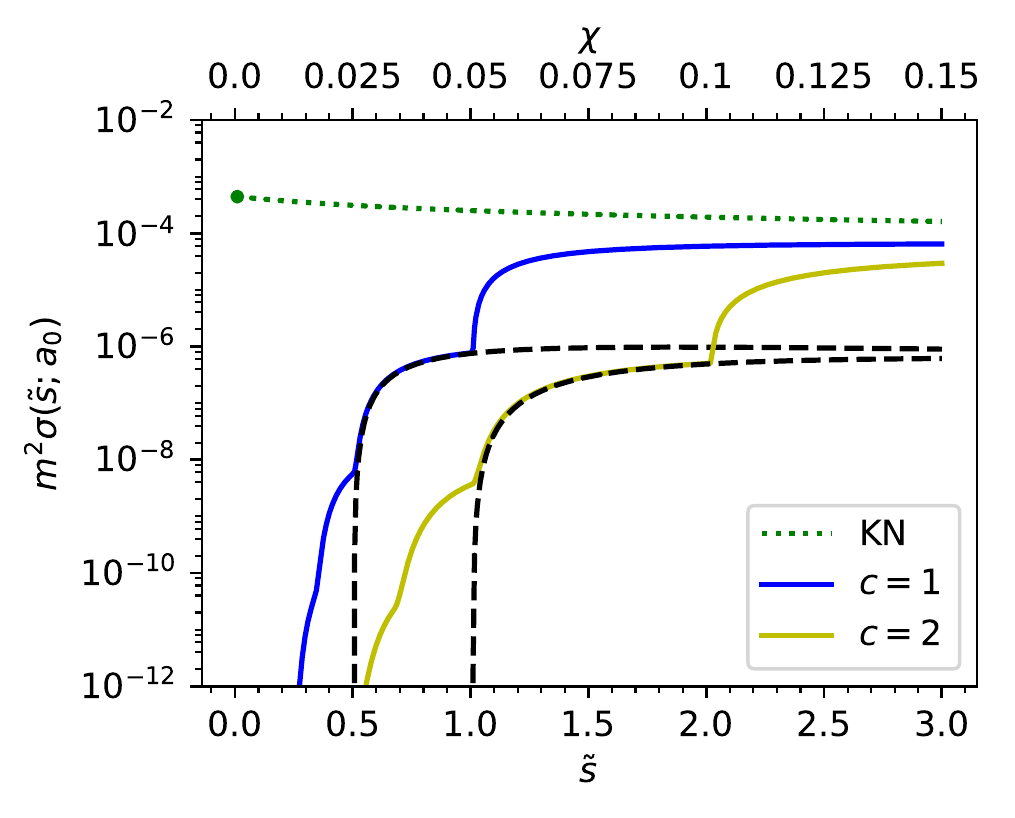}
\put(-410,120){$a_0 = 0.01$}
\put(-170,120){$a_0 = 0.1$}
\vspace*{-3mm}
\caption{(Color online)
Scaled cross section of \nlC (\ref{sigma})
with cut-off values $c= 1$ (blue curves) and $c = 2$ (orange curves)
for $a_0 = 0.01$ (left panel)
and $a_0 = 0.1$ (right panel). The dashed curves depict separately
the harmonic $n = 2$. Note the use of the variables 
$\tilde s \equiv 2 p \cdot k / m^2 = (s_1 - m_*^2) / m^2$ (bottom abscissa)
or $\chi = a_0 \tilde s / 2$ (top abscissa); 
accordingly, the harmonic thresholds are at $\tilde s = c (1 +a_0^2) /n$.
For comparison, the Klein-Nishina cross section 
is depicted by a dotted curve, 
and the Thomson cross section is marked by a circle. 
Upon increasing the value of $a_0$, more and more harmonics are lifted,
and at $a_0 = 1$ (not displayed) already a multitude of harmonics adds up
to generate smooth distributions in the sub-threshold region, i.e.\
below $\tilde s = 1$ (2) for $c =1$ (2). These distributions are discussed 
as functions of $1 / \chi$, instead of $\tilde s$ or $\chi$, 
in section \ref{sect:5} below.
 \label{Fig:mpC}}
 \end{figure}

To highlight this channel closing effect with invariant variables we exhibit
in Fig.~\ref{Fig:mpC} the \nlC cross section (\ref{sigma}) 
as a function of $\tilde s \equiv 2 k \cdot p / m^2$
for $a_0 = 0.01$ and 0.1
for two cut-off values, $c = 1$ and 2. The figure unravels clearly the multi-photon
effects which look completely the same as known from \nlBW,
see figure 3 in \cite{Nousch:2012xe} 
(complementary approaches to multi-photon effects are considered in
\cite{Titov:2019kdk}).  
Thus, the channel closing effect is
exactly analog to sub-threshold \nlBW pair production in region II 
\cite{Titov:2012rd}.
There, the threshold $\hat t = 4$ (depicted as
bullet at bottom of the green top parabola $\hat s \hat u = 1$ in Fig.~\ref{Fig:M}) 
limits the physically admissible
region: only harmonics with $\hat t_n \ge 4$ contribute.
The notion "sub-threshold" means $\hat t_{n=1} < 4$.
Similar to the \nlC process,
we have displayed in Fig.~\ref{Fig:M}
only two possible proxies (horizontal green lines)
of two harmonics of \nlBW in the region II.
Note that, in considering \nlBW pair production {\it per se}, 
one changes usually the coordinate names $\hat t_n \to \hat s_n$
etc.\ according to the crossing symmetry relations \cite{LL}.

\section{Non-perturbative regime, $\mathbf{a_0 \gtrsim 1}$}\label{sect:5}

After enforcing a non-trivial threshold in \nlC process by
the cut-off $c > 0$, one expects a further similarity to the \nlBW
in the region $a_0 > 1$ despite different phase spaces.
As shown originally in \cite{HReiss,Ritus,Nikishov_Ritus}, in the tunneling regime
$a_0 \lesssim 1/\sqrt{\kappa} \gg 1$, the \nlBW pair creation rate scales as
$\propto \kappa \exp\{- 8 / 3 \kappa\}$, where $\kappa = a_0 k \cdot k' / m^2$
(here, $k$ and $k'$ are the $in$ four-momenta of the laser and probe photons). 
In head-on collisions, $\kappa = 2 \frac{\omega'}{m} \frac{E}{E_{crit}}$
since $a_0 = \frac{m}{\omega} \frac{E}{E_{crit}}$.
That yields the Schwinger type dependence
$\propto \exp\{- a_{n \ell BW} E_{crit} / E\}$ with
$a_{n\ell BW} = \frac 43 \frac{m}{\omega'}$.
The large ratio $E_{crit} / E$ can be compensated by a small ratio $m / \omega'$,
thus making the pair creation rate accessible in present day experiments
by using hard probe photons with $\omega' \gg m$,
in contrast to the plain Schwinger rate, even with assistance effects.
As emphasized in \cite{Hartin:2018sha}, such a Schwinger type rate of \nlBW 
is found numerically already for $a_0 \gtrsim 1$ and $\kappa \lesssim 1$.  
   
\begin{figure}[tb!] 
\includegraphics[width=0.49\columnwidth]{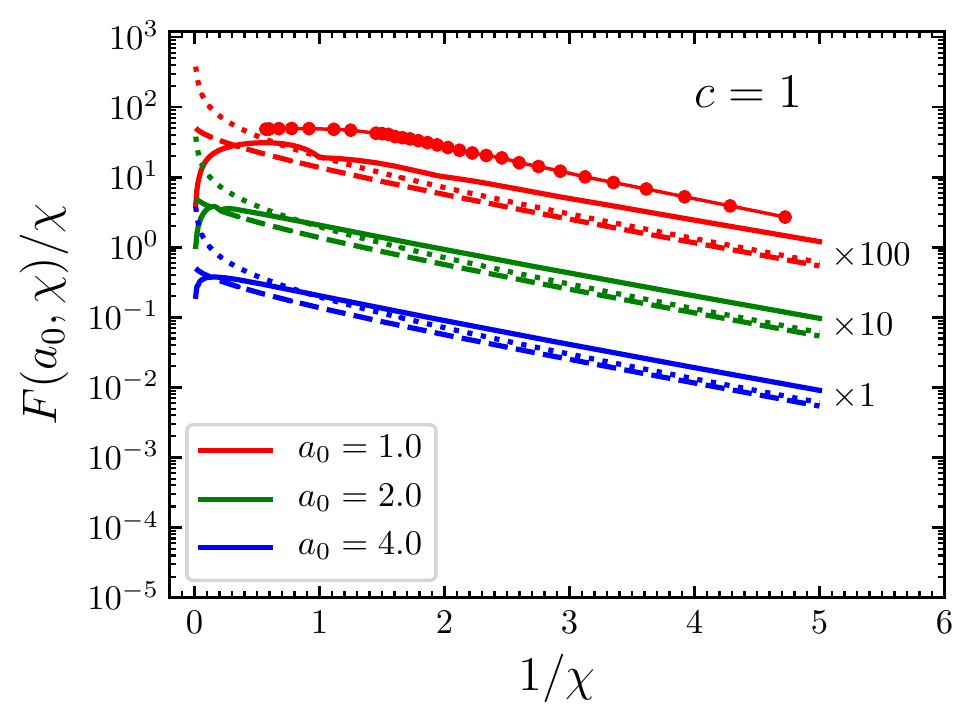}
\includegraphics[width=0.49\columnwidth]{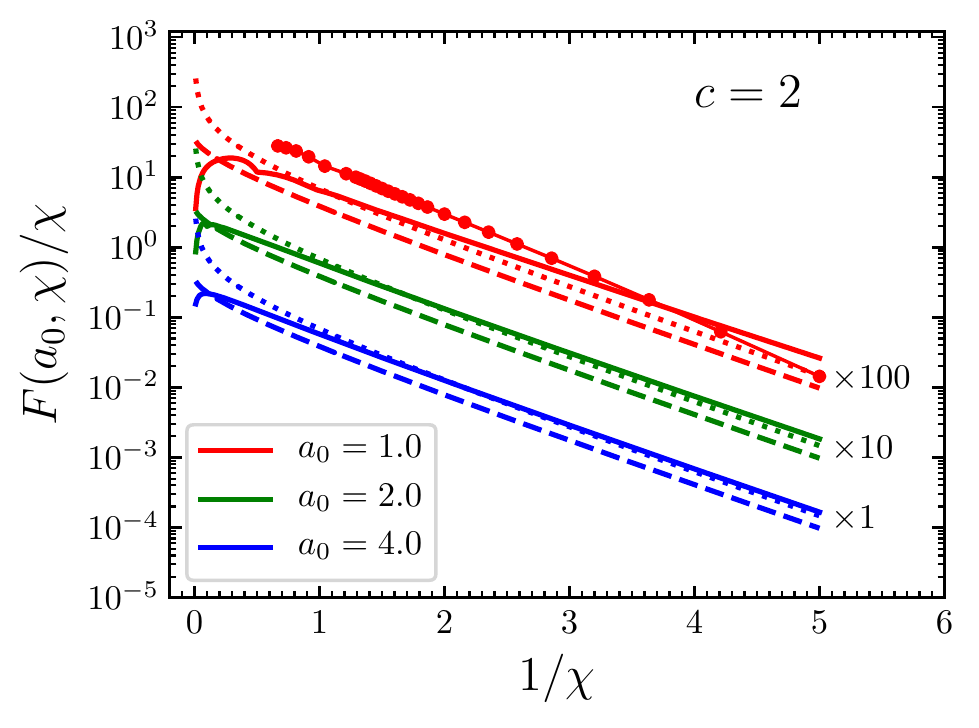}
\vspace*{-4mm}
\caption{(Color online)
Normalized cross section 
$m^2 \sigma(\chi) a_0 / (\alpha^2 \pi) = F(\chi)/\chi $
of \nlC from (\ref{sigma}) as a function of
$1 / \chi$ (solid curves)
with cut-off values $c= 1$ (left panel) and $c = 2$ (right panel).
The dashed (dotted) curves exhibit the approximation (\ref{exp}) with erfc (exp).
For $a_0 = 1$, 2 and 4 from top to bottom with scaling factors as indicated.
The red bullets mark results of QED calculations 
with bandwidth and ponderomotive effects for $a_0 = 1$
(see sub-section \ref{bandwidth}).
Fits of these data by $m^2 \sigma(\chi) \propto \exp\{- 2c_{fit}^c/3\chi \}$
within the interval $1/\chi = (1, 5]$ deliver
$c_{fit}^{c=1} = 1.27$ and $c_{fit}^{c=2} = 2.59$.
\label{Fig:3}}
\end{figure}

Quite in contrast to \nlBW, the \nlC cross section without cut-off
displays a polynomial dependence on the invariant Ritus variable\footnote{
The Ritus variable $\chi$ is a measure of the field strength $E/E_{crit}$ in
the rest frame of the electron; $\chi$ encodes the energy 
of the laser + electron beams as well as the laser intensity. The high-energy
limit and the high-intensity limit do not commute albeit they yield both
a high-$\chi$ asymptotic \cite{Podszus:2018hnz,Ilderton:2019kqp}.} 
$\chi \equiv a_0 k \cdot p / m^2 = a_0 (s_1 - m_*^2) /2 m^2$ \cite{Ritus}. 
However, imposing the cut-off $c > 0$, thus suppressing the low harmonics
in (\ref{sigma}) by a threshold, turns the behavior to an exponential one.
In fact, evaluating (\ref{sigma}) numerically, one obtains the solid curves in Fig.~\ref{Fig:3}
for $c = 1$ (left panel) and 2 (right panel).
Since at $1/\chi < 1$ the curves display an $a_0$ dependence, we
have employed scaling factors. Without the latter ones, the curves
at $1 / \chi > 1$ are nearly perfectly on top of each other, i.e.\
independent of $a_0$. To quantify the $1 / \chi$ dependence we depict
for a comparison the dashed/dotted curves based on
\begin{equation} \label{exp}
F_\infty(\chi, c) = \sqrt\frac32 \frac{\chi}{\pi} f(c) \, {\rm erfc} 
\left( \sqrt{\displaystyle\frac{2 c}{3 \chi}} \right) + {\cal O}(\chi^{2/3})
 \stackrel{c \gg \chi}{\longrightarrow}
\left( \frac{\chi}{\pi} \right)^{3/2} f(c) 
\exp\left\{ - \displaystyle\frac{2 c}{3 \chi} \right\} + \cdots,
\end{equation} 
where $f(c) = (5 + 7 c +5 c^2)/(1+c)^3$ and
``erfc" stands for the complementary error function. 

One avenue to (\ref{exp}) is to start with (\ref{sigma})
in the limit $a_0\to \infty$ with side condition 
$(1 -z_n^2 / n^2) a_0^2 = const$ and then to convert the sum
via the Euler-Maclaurin formula into an integral,
$F_\infty (\chi, c) = - \frac{4}{3 \pi}
\int_c^\infty dx \, f(x) x^{-2/3} Ai' (z(x))$
with $z(x) = (x/\chi)^{2/3}$.
Under the condition $c \gg \chi$,
the derivative of the Airy function, $Ai'$, can be replaced by
its asymptotic representation and the integral
can be executed upon a shift of the variable $x$ and a suitable Taylor expansion.
 
Surprisingly, the small-$\chi$ leading-order term 
$\propto \exp \{ - 2 c / 3 \chi \}$ in (\ref{exp}) numerically 
approximates (\ref{sigma}) fairly well in the non-asymptotic region,
$a_0 \gtrsim 1$ and $\chi < 1$, irrespectively of the assumptions made
in the sketched derivation.
As a consequence, the
\nlC cross section also displays a Schwinger type
dependence $\sigma(c > 0) \propto \exp\{ - a_{n \ell C} E_{crit}/E \}$
for suitable values of the cut-off $c > 0$, in general with $a_{n \ell C}(c, a_0, s_1)$.
That is, the paradigmatic transmonomial behavior 
\cite{Aniceto:2018bis}
is provided not only for pair creation but shows up also in high-harmonics Compton scattering on the level of ``total" cross section,
which actually means integration over a fraction of the $out$-phase space. 

\section{Discussion}\label{sect:6}

\subsection{Imposing the cut-off: kinematics}

The cut-off $c > 0 $ in (\ref{sigma}) looks quite innocent, but in practice it may
become challenging. To illustrate that feature let us employ laboratory 
observables: $\nu \equiv \omega /m (= 2 \times 10^{-6}$ for optical lasers),
$p_0 / m = \cosh \zeta$ the Lorentz factor of the $in$-electron, 
$\nu' \equiv \omega' /m$ for the normalized energy of the $out$-photon
in direction $\Theta'$ such that $\Theta' = 0$ and $\Theta' = \pi$ mean
on-axis forward scattering and on-axis backscattering, respectively. Adopting the 
notation in \cite{Harvey:2009ry} we recall the relation
\begin{equation} \label{x_eq}
x = \frac{(1 - \cos \Theta') \nu'}{e^\zeta -(1 - \cos \Theta') \nu'}.
\end{equation}
The admissible intervals for a harmonic $n$ are for head-on collisions
$0 \le x \le y_n \equiv 2 n \nu e^\zeta / (1 + a_0^2)$,
$n \nu  \le \nu_n' \le  n\nu /(1 + 2 \kappa_n e^{-\zeta}) $
for $2 \kappa_n \equiv  2 n \nu - e^{\zeta} + (1 + a_0^2) e^{-\zeta} < 0$
or $  n \nu  /(1 + 2 \kappa_n e^{-\zeta}) \le \nu_n' \le  n \nu $ for $\kappa_n > 0$,
and $0 \le \Theta' \le \pi$. 
One has also to recall the well known \nlC kinematic relation
$\nu' (n, \Theta'; a_0, \nu, \zeta) = n \nu /\left[1 + \kappa_n e^{- \zeta} (1 - \cos \Theta') \right]$,
e.g.\ in relating the $x$ and $\Theta'$ coordinates:
a point at $x = \xi y_n$ corresponds to 
$(\pi - \Theta')^2 \approx 4 \frac{1-\xi}{\xi} (1 + a_0^2) e^{-2 \zeta}$,
independent of the harmonic number. This highlights the preference
of backscattering in the relativistic regime, since a significant fraction
of events with $x \to y_n$ is seen at $\Theta' \to \pi$.     

\begin{figure}[t] 
\includegraphics[width=0.55\columnwidth]{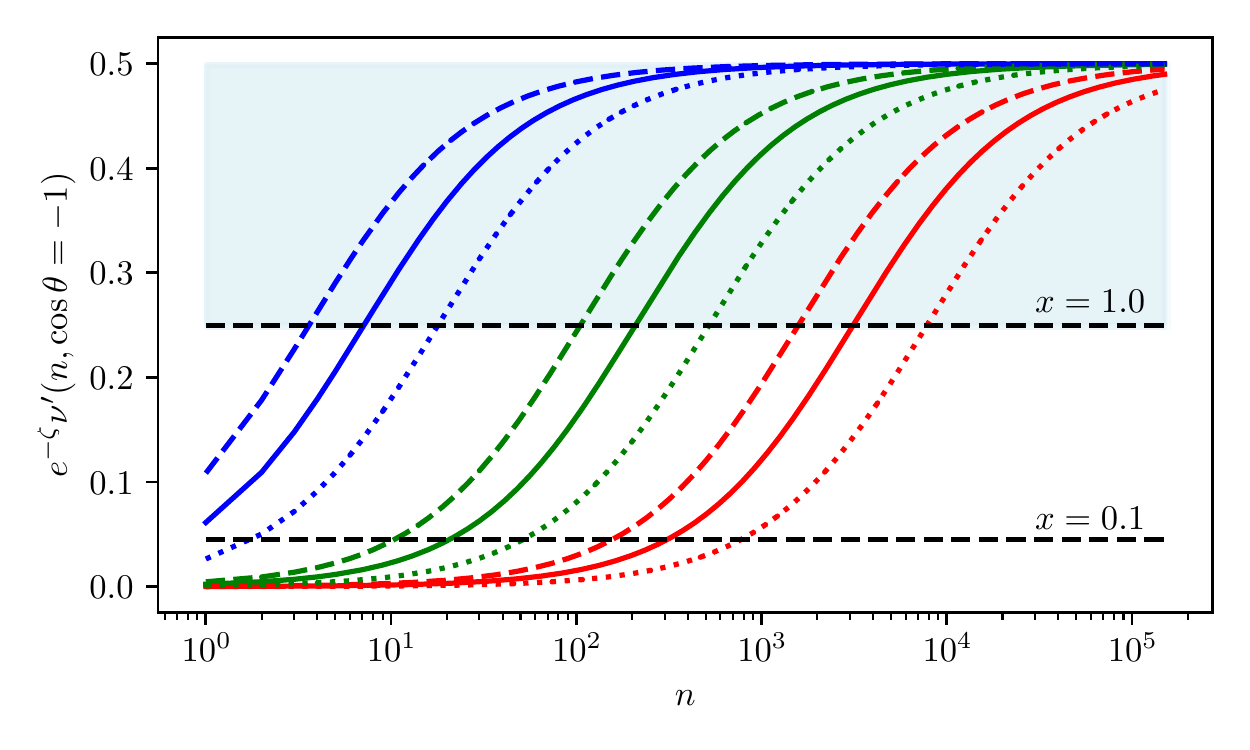}
\includegraphics[width=0.44\columnwidth]{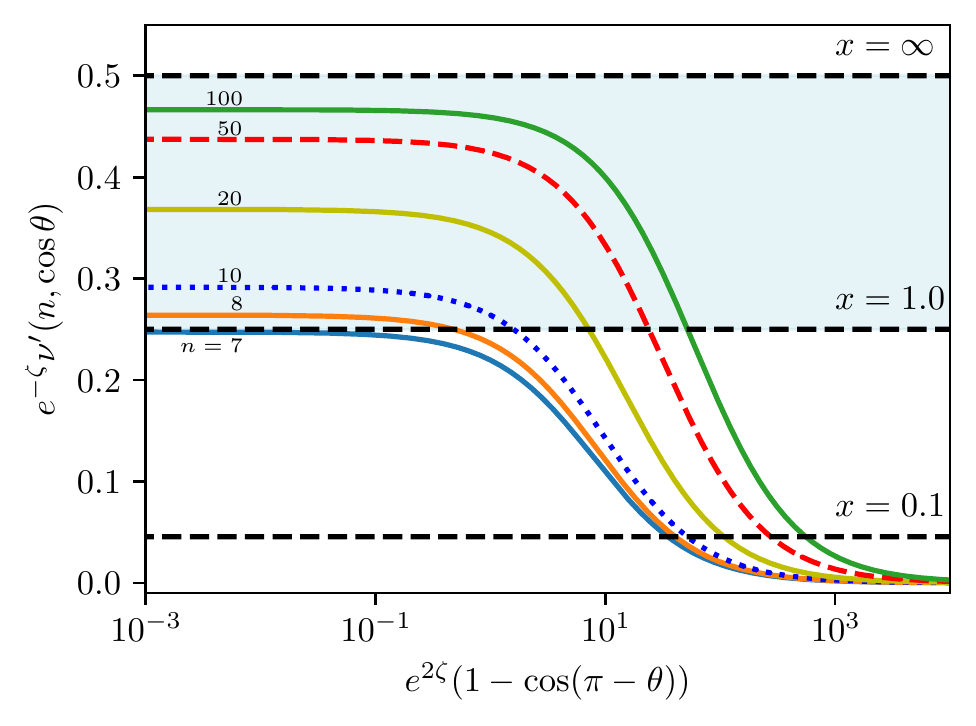}
\vspace*{-15mm}
\caption{(Color online)
Left panel: The scaled $out$-photon energy $\nu'$ as a function of the harmonic
number $n$ for three kinematic situations referring to the entrance channels
at LUXE (left, in blue), ELI (middle, in green), and HZDR (right, in red).
For $a_0 = 0.01$ (dashed), 1 (solid) and 2 (dotted) and on-axis backscattering.  
The curves connect smoothly the values of $\nu'$ at the discrete harmonic
numbers $n$.
Selecting the $out$-channel with $x \ge c$ means 
accepting only events in the light-blue region if $c = 1$ is chosen. 
Right panel: The scaled $out$-photon energy as a function of the
angle $\Theta'$ for the harmonics $n = 7$, (black thin), 
10 (blue dotted), 20 (red solid), 50 (green dashed) and 100 (black thin)
(LUXE parameters).
Note the relation $1 - \cos (\pi -\Theta') \approx\frac12 \vartheta'^2$
for backscattering, where $\vartheta' = \pi - \Theta'$ measures the angle of the
$out$-photon relative to the $in$-electron direction. 
The seemingly horizontal dashed lines 
$e^{- \zeta} \nu' = \frac{x}{1+x} \frac{1}{1 - \cos \Theta'}$
are depicted for three values of the invariant
quantity $x =0.1$, 1 and $\infty$, and the light-blue region is again for admissible
events if $c = 1$ is chosen.
The harmonic $n = 7$ does never enter the light-blue region.
\label{Fig:nu_prime}}
 \end{figure}

These relations evidence that one has to
reject events with too low values of $\nu'$ or select sufficiently high harmonics
to realize the request $x \ge c$, see left panel of Fig.~\ref{Fig:nu_prime}
for on-axis backscattering. 
The meaning of these curves is that the realization of $x \ge c$ requires in general
$\nu' (n, \Theta'; a_0, \nu, \zeta) \ge \nu'(x, \Theta'; \zeta)$ 
either as a function of $n$ (left panel, for $\Theta' = \pi$) or 
as a function of $\Theta'$ (right panel, for selected harmonics),
where (\ref{x_eq}) determines the 
$\nu$ independent function $\nu'(x,\Theta'; \zeta)$.
These relations are exhibited also in the right panel of Fig.~\ref{Fig:nu_prime},
where the light-blue region depicts the range wherein $x \ge c = 1$ is fulfilled. 
In the preferred backward direction $\Theta' \to \pi$, 
the curves $\nu'(x,\Theta'; \zeta)$ (black dashed)
are nearly horizontal, with the benefit that only an energy-resolved measurement
is necessary to select the wanted range $x \ge c$ by imposing a veto for
all events with $e^{2 \zeta} ( 1 - \cos(\pi - \Theta')) > 20$, for instance.
At smaller angles $\Theta'$,
i.e.\ going further to the right -- beyond the region displayed in the right panel of Fig.~\ref{Fig:nu_prime} --
the curves $\nu'(x,\Theta'; \zeta)$ bend up, 
which would require also an angular-resolved
measurement. However, the contributions of the very high harmonics are
exceedingly small in that phase space region and can be neglected.
Altogether, a cross section measurement in backward direction
and above the threshold value
$e^{ \zeta} \nu' \ge 0.25$ facilitates the cut-off $c = 1$ for the LUXE kinematics.  

\subsection{Dead cone}

\begin{figure}[tb] 
\includegraphics[width=0.48\columnwidth]{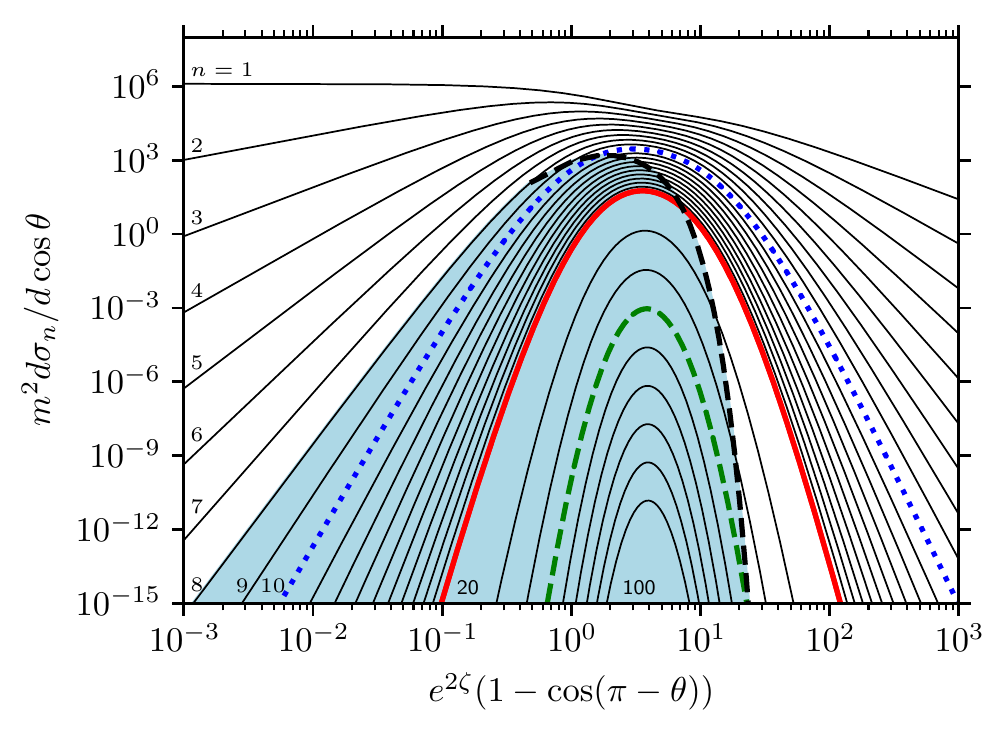}
\includegraphics[width=0.45\columnwidth]{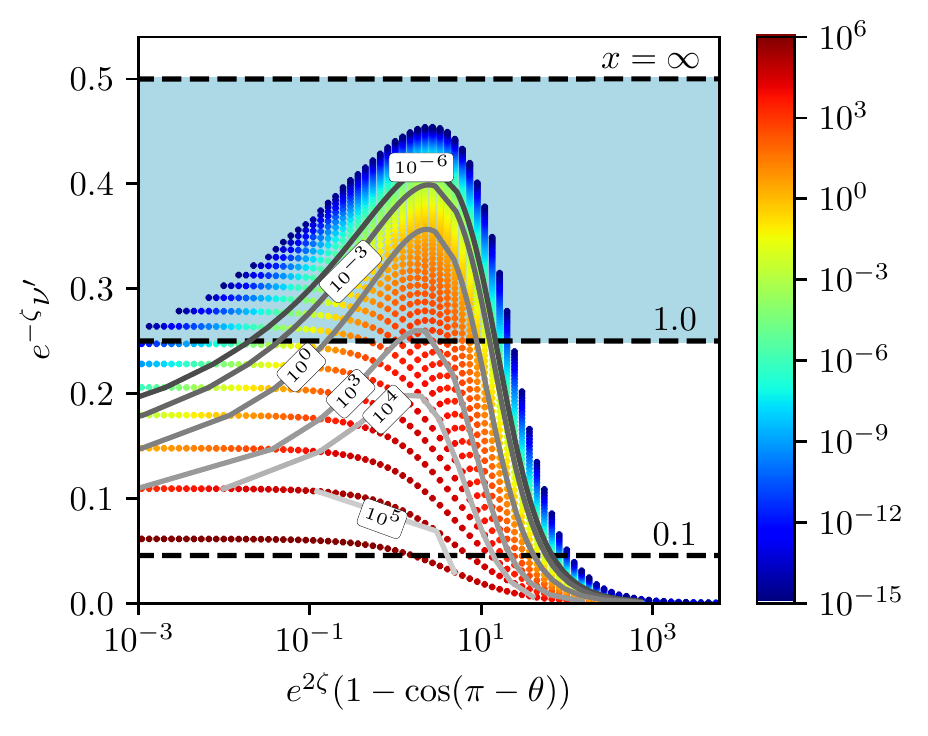}
\vspace*{-3mm}
\caption{(Color online)
Left panel:
Angular differential cross sections 
$m^2 d \sigma_n / d \cos \Theta' = 
\frac{\alpha^2 \pi}{a_0^2}
\frac{m^2}{k \cdot p}
\frac{e^\zeta}{n \nu (1  - \cos \Theta')^2}
\frac{x^2}{(1+x)^2}
F_n(z_n)$
with $x$ and $z_n$ to be viewed as functions of $\Theta'$
(cf.\ \cite{Harvey:2009ry})
for the harmonics $n = 1 \cdots 20$ (in steps of 1) and $20 \cdots 100$
(in steps of 10). The harmonics $n = 10$ (blue dotted), 20 (read solid)
and 50 (green dashed) are depicted in color code and line style as in 
Fig.~\ref{Fig:nu_prime}-right; the other harmonics (black solid) are partially
labeled. The black dashed curve -- to be continued by the harmonic
$n = 8$ -- limits the admissible range (in light-blue)
where $x \ge c = 1$ is fulfilled.
Right panel: Angular differential cross sections $m^2 d \sigma_n / d \cos \Theta'$ 
(see color code on r.h.s.)
over the scaled $\nu'$ - $\Theta'$ plane. 
Low-order harmonics are clearly separated
for monochromatic lasers (but are smeared out for pulses, see below).
The gray lines connect points of equal values 
(given by the numbers in the boxes)
of $m^2 d \sigma_n / d \cos \Theta'$
on adjacent harmonics.  
Contributions smaller than $10^{-15}$ are not displayed. 
Both panels are for the LUXE kinematics with 
$\nu = 2 \times 10^{-6}$, $e^\zeta = 7 \times 10^4$ and  $a_0 = 1$.
\label{fig:6}}
\end{figure}

In addition to this purely kinematic relations one has to account for the dynamics,
in particular the dead cone effect which is special for circularly polarized lasers 
according to (\ref{sigma}): Ignoring for the moment being the cut-off,
the harmonics $n > 1$ are (multiply) peaked within the interval
$0 < x < y_n$ and drop smoothly towards zero at the boundaries $x \to 0$ and
$x \to y_n$. Transforming to the laboratory frame, the angular differential
cross sections $d \sigma_n / d \cos \Theta'$ of selected harmonics $n > 1$
are peaked as exhibited in Fig.~\ref{fig:6}-left. Only the $n = 1$ harmonic remains
non-zero for $\Theta' \to \pi$. The dropping of 
$d \sigma_n / d \cos \Theta' \vert_{n > 1}$ at the left side is the dead cone effect.
The dropping at the right side refers to the suppression of forward scattering,
i.e.\ at $\Theta' \to 0$. The black dashed curve connects the points of intersections
of the curves $\nu' (n, \Theta'; a_0, \nu, \zeta)$ and $\nu' (x=1, \Theta'; \zeta)$,
which can be read off in Fig.~\ref{Fig:nu_prime}-right for $n = 10$, 20, 50
and 100. For $n < 8$, there are no such intersections and, as a consequence,
the harmonic $n = 8$ -- left-beyond the dashed curve -- limits the admissible
region $x \ge c = 1$ which is highlighted in light-blue, as in Fig.~\ref{Fig:nu_prime}.

As advertised above, the contributions of the high harmonics at 
$e^{2 \zeta} ( 1 - \cos(\pi - \Theta')) > 20$ are exceedingly small, thus
substantiating our claim that imposing a frequency 
threshold is enough for a measurement of $\sigma(c = 1)$,
at least for the here employed LUXE kinematics. This is evidenced in Fig.~\ref{fig:6}-right,
where some proxy of a contour plot (gray lines) of the angular differential cross sections 
$m^2 d \sigma_n / d \cos \Theta'$ 
is exhibited over the scaled $\nu'$ - $\Theta'$ plane. [The harmonics $n$ have only 
support on the curves  $\nu' (n, \Theta'; a_0, \nu, \zeta)$ (see  Fig.~\ref{Fig:nu_prime}-right
for several values of $n$), and the colored pixels encode the values of $m^2 d \sigma_n / d \cos \Theta'$.
The gray lines connect points of equal strengths on adjacent harmonics, thus serving as contour
lines despite the discrete occupancy in $\nu'$ direction.]
The experimental challenge is therefore
the precise setting of a frequency threshold to select $x \ge c$ events out of a large background. 

\subsection{Bandwidth effects and ponderomotive broadening}\label{bandwidth}

\begin{figure}[tb!] 
\includegraphics[width=0.49\columnwidth]{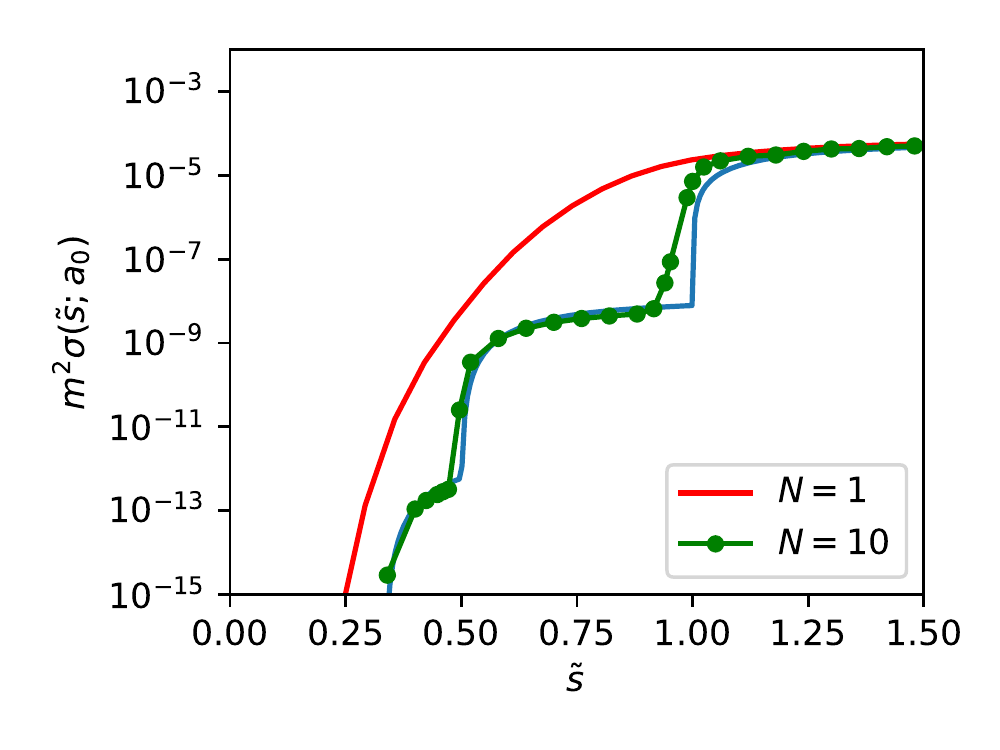}
\includegraphics[width=0.49\columnwidth]{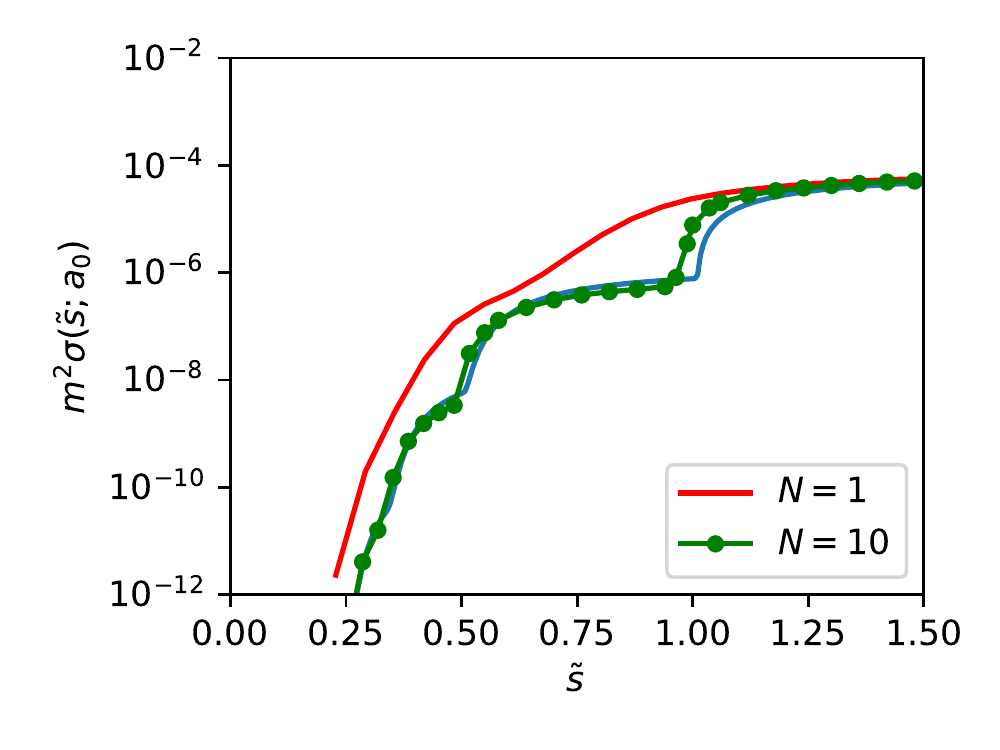}
\put(-395,140){$c = 1$}
\put(-155,140){$c = 1$}
\put(-395,130){$a_0 = 0.01$}
\put(-155,130){$a_0 = 0.1$}
\vspace*{-6mm}
\caption{(Color online)
Same as in Fig.~\ref{Fig:mpC} but for  
laser pulses with envelope shape $g(\phi) = 1/\cosh(\phi/N \pi)$.
$N =1$ (red curves) represents ultra-short pulses  
and $N = 10$ (green curves marked by dots) is among currently available
short pulses.   
Blue curves are as in Fig.~\ref{Fig:mpC} for the monochromatic case.
For the cut-off value $c = 1$ and $a_0 = 0.01$ (left panel) 
and 0.1 (right panel). 
 \label{Fig:mpC_FPA}}
 \end{figure}

While (\ref{sigma}) is for monochromatic laser beams
with the four-potential
$\vec A(\phi) = g(\phi) \left[ \vec a_1 \cos \phi + \vec a_2 \sin \phi \right]$
of the e.m.\ field with invariant phase $\phi$ and
obeying $g(\phi) = 1$, $\vec a_1 \vec a_2 = 0$, $\vec a_1{}^2 = \vec a_2{}^2$,
one has to check whether laser pulses\footnote{
The Fourier-zero mode of the non-linear phase can not be longer absorbed
in a redefinition of the electron momentum as quasi-momentum 
\cite{Narozhnyi:1996qf}, 
which is a key quantity in the monochromatic laser beam model (\ref{sigma}),
but an expansion into harmonics is still possible for smooth, long pulses
\cite{Narozhnyi:1996qf,Heinzl:2020ynb}
(cf.\ \cite{SeiptDaniel:2012tua} for a discussion of these issues and a 
compendium of one- and two-photon emission off electrons in laser pulses).} 
are well approximated when 
focusing on total cross sections.    
In Fig.~\ref{Fig:mpC_FPA}, the cross section
as a function of $\tilde s$ is exhibited,
for $a_0 = 0.01$ and 0.1 as in Fig.~\ref{Fig:mpC},
however, for short and ultra-short pulses.
The QED calculations are based on equations (40, 42) in \cite{Titov:2019kdk} 
(version v1) with $u \ge c$ to impose the cut-off. 
The pulse shape envelope is here especially
$g(\phi) = 1 /\cosh (\phi / N \pi)$, where $N$ characterizes the number of 
oscillations of the field. This envelope $g(\phi)$
does have neither an extended flat-top section 
nor narrow ramping sections. The former property makes it quite different to a
near-monochromatic beam with broad flat-top envelope
and may be considered as representing a ``worst case" in that respect.
The related bandwidth effects and longitudinal ponderomotive broadening
are fully included in the QED calculation of one-photon emission in
\cite{Titov:2019kdk}. For short pulses, these effects
smoothen the step like shape of the
total cross sections, as known from \nlBW  \cite{Nousch:2012xe}.
In particular, for the ultra-short pulse with $N = 1$ (red curves), 
the combined strong bandwidth effect and ponderomotive broadening
overwrite completely the multi-photon effects; the cross section
is stark enhanced in the sub-threshold region.
However, for sufficiently long pulses with $N \ge 10$, 
i.e.\ a pulse duration of $> 30$ fs for optical laser pulses, 
even without pronounced temporal flat-top profile,
the essential dependencies of the
cross section model with cut-off (\ref{sigma}) are recovered,
see green curves marked by dots in Fig.~\ref{Fig:mpC_FPA} for $a_0 = 0.01$ and 0.1
and red curves marked by bullets in Fig.~\ref{Fig:3} for $a_0 = 1$.
Since the normalization of cross sections in pulses is special 
(cf.\ \cite{Titov:2019kdk,Titov:2014usa}), let us focus on slopes at $a_0= 1$.
As noted in the caption of Fig.~\ref{Fig:3}, 
the slope parameters $c_{fit}^c$ of the pulse
model with $N = 10$ are about 25\% larger than the ones of the monochromatic
model (\ref{sigma}). Despite these differences, the ratio is still $1 : 2$.
Such a cut-off dependence can be experimentally tested in the analysis
of a given data set after data taking, e.g.\ at a suitable value of $\chi$.
(A comprehensive theoretical study of the $a_0 > 1$ dependence must be postponed
because our present numerical implementation restricts us to $a_0 \le 1$.) 

\begin{figure}[tb] 
\includegraphics[width=0.47\columnwidth]{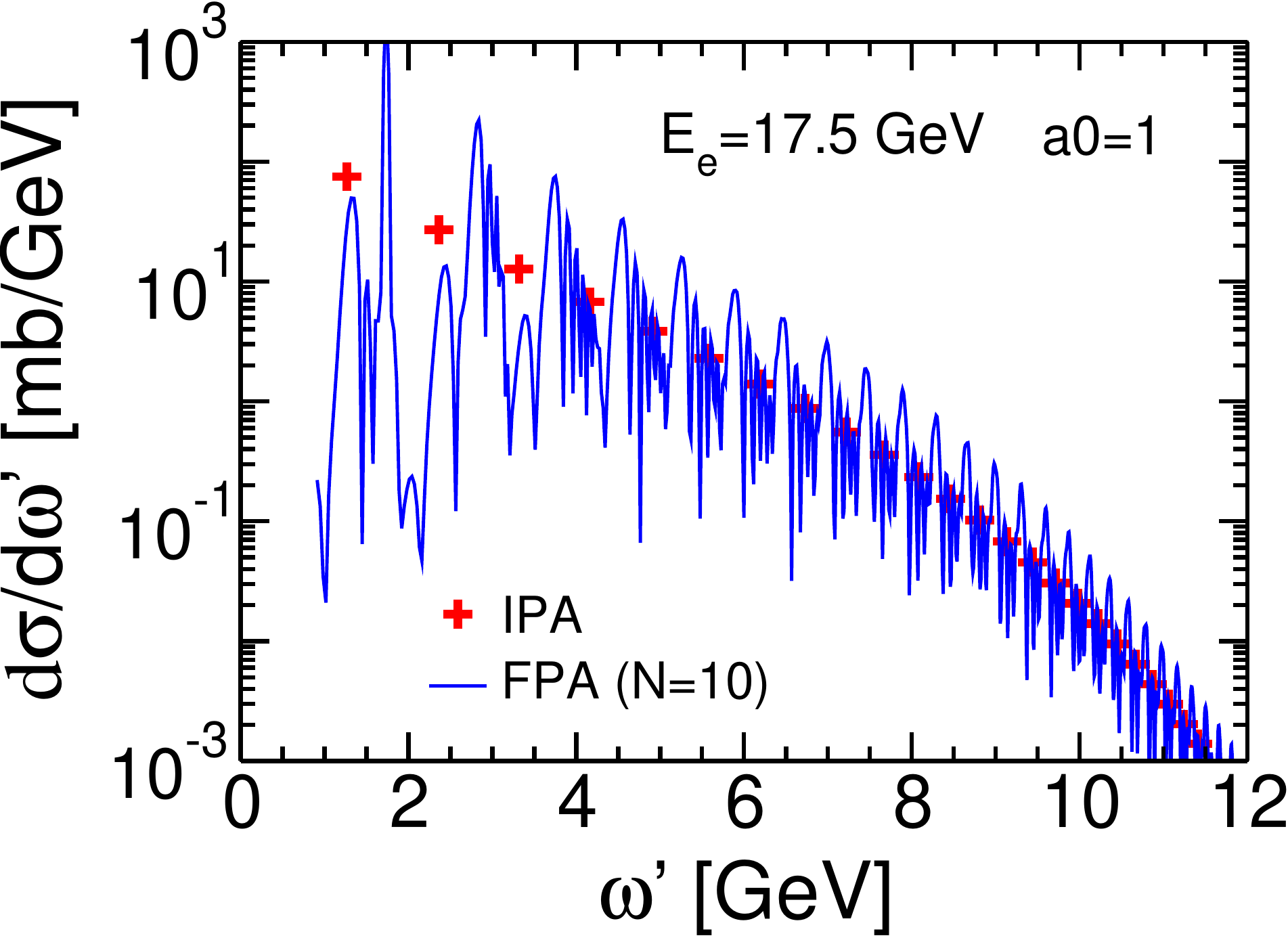}
\quad
\includegraphics[width=0.47\columnwidth]{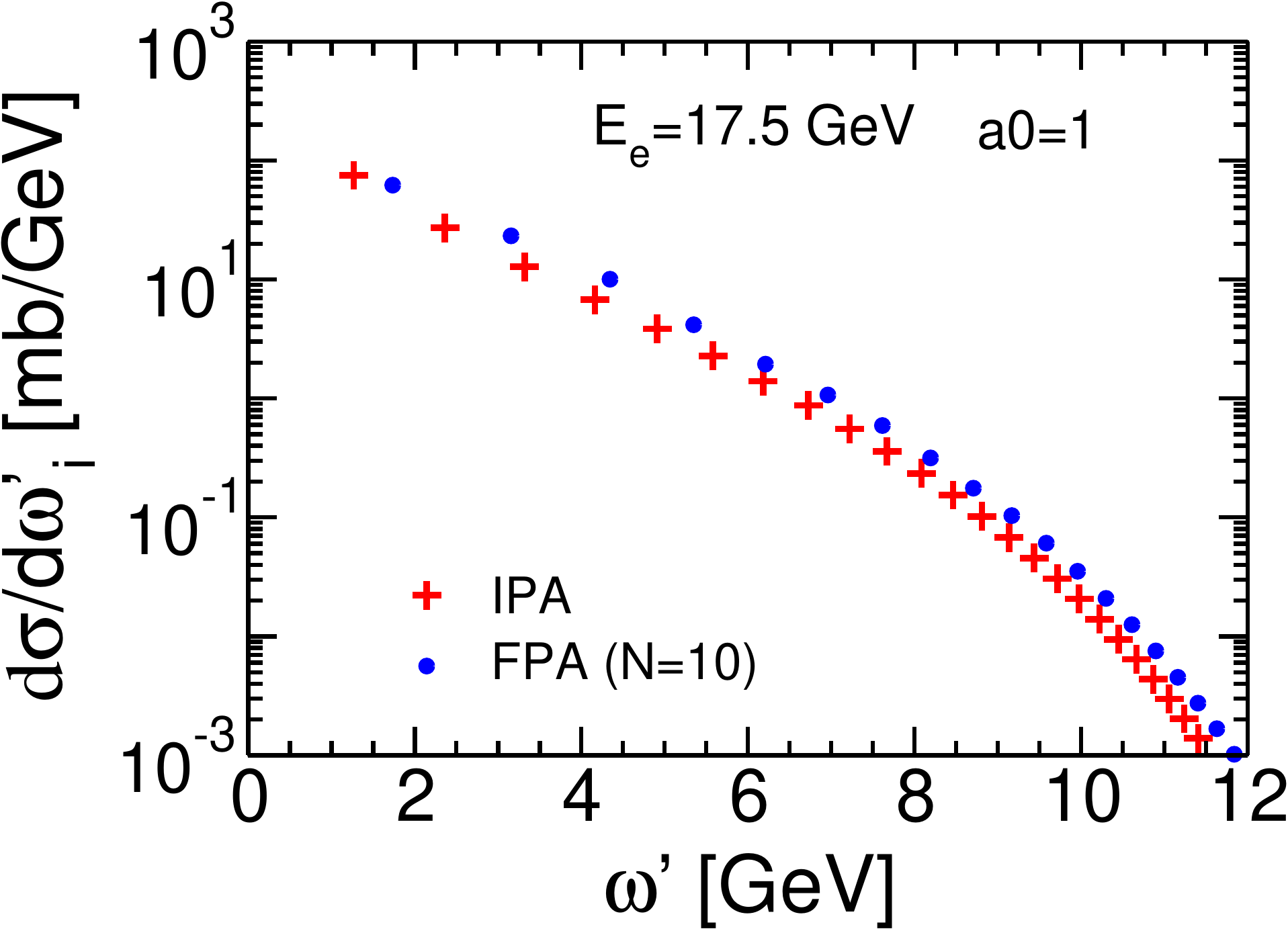}
\vspace*{-3mm}
\caption{(Color online)
Left panel: 
The differential spectrum $d \sigma / d \omega' \vert_{\Theta'}$ 
as a function of $\omega'$ at polar angle $\Theta'$ determined by 
$e^{2 \zeta} (1 - \cos (\pi - \Theta')) = 3$
for the $1/\cosh (\phi/ N \pi)$ pulse envelope with $N = 10$ (blue curve).
The corresponding QED calculation is as in figure 5 of \cite{Titov:2014usa}
but for the LUXE kinematic parameters used in Fig.~\ref{fig:6}.
Red pluses mark the peak positions in the monochromatic model (\ref{sigma}).
Right panel:
The same as in the left panel but with locally averaged cross section (blue bullets)
over the intervals $\omega' (\ell - 0.5) \cdots \omega'(\ell + 0.5)$
for the internal auxiliary variable $\ell = 1, 2, 3 \cdots$
(cf.\ equation (16) in \cite{Titov:2014usa}). The legends adopt the nomenclature
in \cite{Titov:2014usa}: IPA stands for the monochromatic laser beam and FPA
denotes the laser pulse model.
\label{fig:broadening}
}
\end{figure}

Turning to details of differential observables, the model (\ref{sigma}) 
is in general a less useful guide. 
In fact, keeping the above pulse shape parameterization
by $g(\phi) = 1/ \cosh(\phi/N \pi)$, the differential cross section
$d \sigma / d \omega' \vert_{\Theta'}$, e.g.\ 
for $e^{2 \zeta} (1 - \cos(\pi - \Theta')) = 3$ (that is for $\Theta'$ at
about the maximum of the angular differential cross section in
Fig.~\ref{fig:6}-left), does hardly recover the harmonic structures which
can be deduced from
the right panels of Figs.~\ref{Fig:nu_prime} and \ref{fig:6},
even for longer pulses, see left panel in Fig.~\ref{fig:broadening}.
Instead of clearly recognizable peaks at 
$\omega' = m \nu'(n, \Theta'; a_0, \nu, \zeta)$, 
the spectrum in Fig.~\ref{fig:broadening}-left
displays some complexity which is further obscured by increasing gradually the
parameter $N$. This feature is known since some time,
cf.\ \cite{Narozhnyi:1996qf,Titov:2014usa,Seipt:2010ya} for instance. 
Bandwidth effects and ponderomotive broadening have been identified as responsible,
together with interferences.

To highlight the impact of the former ones
it is instructive to cast the above
\nlC kinematic relation $\nu' (n, \Theta'; a_0, \nu, \zeta)$ in the form
\begin{equation} \label{eq:broadening}
2 e^{- \zeta} \nu' = 
\frac{1}{1 + \frac{e^{-\zeta}}{2 n \nu \mu} (1 + a_0^2 (\phi))
+ e^{2 \zeta} (1 - \cos(\pi - \Theta'))
\frac{e^{- \zeta}}{4 n \nu \mu} \left[
1 - \frac{2 n \nu \mu}{e^\zeta} - e^{-2 \zeta} (1 + a_0^2 (\phi)) \right] } ,
\end{equation}
where $a_0 (\phi)$ puts emphasis on the longitudinal ponderomotive broadening
by the variation of the intensity in the course of a pulse, $0 < a_0(\phi) \le a_0$, 
and $\mu \ne 1$ accounts for the bandwidth effects, i.e.\ there is a distribution
of laser frequencies around the central frequency $\nu$.  
These effects are seen best in very backward kinematics, where
$e^{2 \zeta} (1 - \cos(\pi - \Theta')) \to 0$: The support of harmonic $n$
is in the interval 
$\left( 1 + \frac{e^{-\zeta}}{2 n \nu \mu} (1 + a_0^2) \right)^{-1}
\le 2 e^{- \zeta} \nu' \le
\left( 1 + \frac{e^{-\zeta}}{2 n \nu \mu} \right)^{-1}$
and depends additionally on the frequency spread parameterized by $\mu$.
The net effect is blowing up the curves $\nu'(n, \Theta')$ in 
the right panels of Figs.~\ref{Fig:nu_prime} and \ref{fig:6}
to overlapping bands (not displayed), 
already either for $a_0 \ge 1$ or $0.5 < \mu < 2$ separately.
Both effects, $a_0(\phi)$ and $\mu \ne 1$, can be separated only
in certain asymptotic regions 
\cite{Titov:2014usa}. Besides rich sub-structures within the broadened
and overlapping harmonic support regions,
QED calculations for laser pulses exhibit complicated
interference patterns over the $\omega'$-$\Theta'$ plane
depending on the actual pulse envelope shape $g(\phi)$ and its parameters,
see figure 1 in \cite{Seipt:2011dx} for an example.

Nevertheless, the model (\ref{sigma}) can provide a useful guide for the
gross shape of the locally averaged spectrum. Averaging the differential
cross section, exhibited in Fig.~\ref{fig:broadening}-left by the blue curve,
over the intervals $\omega' (\ell - 0.5,\Theta') \cdots \omega'(\ell + 0.5, \Theta')$
for $\ell = 1, 2, 3 \cdots$\footnote{\nopagebreak
The continuous variable $\ell$ is an internal auxiliary quantity
which replaces the harmonic number $n$ in the case of a pulse with
smoothly varying envelope (cf.\ equations (16, 17) in \cite{Titov:2014usa}).}
leads to the spectral shape displayed by
blue bullets in the right panel of Fig.~\ref{fig:broadening}.
Within a factor of two, both spectral shapes 
-- the one based on the monochromatic model (\ref{sigma}) and
the one with $1 / \cosh$ pulse envelope -- 
agree over six orders of magnitude.
The displacement of pair-wise related red crosses and blue bullets in
the right panel of Fig.~\ref{fig:broadening} can be attributed to the frequency
difference in (\ref{eq:broadening}) for $\nu' (a_0)$ and $\nu'(a_0=0)$
at the same value of $n$ when ignoring the bandwidth effect.

If one wishes to recover the clear harmonic structures exhibited in
Fig.~\ref{fig:6}-right, one has to employ suitable laser pulse shapes,
e.g.\ with extended flat-top profiles (see figure 6 in \cite{Seipt:2010ya})
or
frequency chirping \cite{Seipt:2014yga} etc.
Nevertheless, we argue that these details, which shape the differential spectra,
have a sub-leading impact on the phase space integrated cross section, and the
prediction in Fig.~\ref{Fig:3} is essentially robust
within the range uncovered by our pulsed QED calculations 
(red curves marked by bullets) 
and the model (\ref{sigma}) (red solid curves).

In addition to such effects, there is
transverse broadening w.r.t.\ to multiple photon emission: The incoming
electron may suffer a (or many) transverse kick(s) due to soft-photon emission prior
to hard-photon emission, thus not being longer subject of a head-on collision.
Since for the above LUXE kinematics our focus
is on the hard-photon tail, e.g.\ with $\omega' > \frac12 p_0$ for $c = 1$, we do not
expect a significant impact of the leakage of low harmonics into
this region and multiple photon emission and radiation reaction as well. 
For a proper quantitative account, the simulation tools
developed in view of the recent experiments \cite{Poder:2018ifi,Cole:2017zca}
should be employed in dedicated analyses
and compared with analog QED calculations. 

\section{summary}\label{sect:7}

In summary we point out that the non-linear Compton process
obeys a field strength dependence 
$\propto {\cal P} \exp\{- a_{n\ell C} E_{crit} /E\}$,
similar to the Schwinger process of ``vacuum break down", when
imposing a suitable cut-off $c$ which suppresses the low harmonics.
We focus on the slope coefficient 
$a_{n \ell C} =\frac23 c m /(p_0 + \sqrt{p_0^2 -m^2)}$ by a comparison with
some approximation formula which displays a dependence
$\propto \exp\{ - 2 c / 3 \chi \}$ already in the non-asymptotic region. 
Albeit the Compton process does not have such an obvious tunneling regime
as the pair production processes, its formal similarity with the
non-linear Breit-Wheeler process  provides evidence \cite{Ritus,Dinu:2018efz}
for selected {\sl differential} contributions with an exponential field dependence.
The here introduced cut-off acts as a threshold
and enforces a large gap between $in$- and $out$-Zel'dovich levels
(which suffer some broadening in laser pulses) 
or, equivalently, a large light-cone momentum-transfer
from the $in$-electron to the $out$-photon; 
it makes the otherwise
hidden exponential contributions visible in the {\sl ``total"} cross section,
which actually refers to a fraction of the $out$-phase space.
This opens another
avenue towards a measurement of the boiling point of the vacuum,
complementary to plans of the LUXE collaboration by utilizing the
non-linear Breit-Wheeler pair production \cite{Hartin:2018sha}. 
While fort the latter one a high-energy photon beam is vital, 
our approach requires either a moderately high-energy ($p_0$) electron beam 
and the selection of very high harmonics 
or a high-energy electron beam and the selection of moderately high harmonics.    
The experimental challenge is anyway the isolation of the high harmonics region
characterized by the $out$-photon kinematics.

The present considerations apply primarily to a plane-wave, 
monochromatic laser beam, i.e.\
a long flat-top pulse duration, with circular polarization. 
Selected examples of one-photon emission 
in laser pulses, based on Furry picture QED calculations of the cross section,
support such a clear-cut approach.
Nevertheless, necessary obvious extensions
should take into account general laser polarizations as well as
further bandwidth effects, ponderomotive broadening and
multiple photon emissions in
finite-duration pulses and their detailed temporal structures
together with a larger range of the laser intensity parameter $a_0$.
Planned follow-up work is devoted to energy- and angular-differential
spectra and suitable realizations of the crucial cut-off implementation
in non-perfect head-on collisions.   
 
{\bf Acknowledgments:}
The authors gratefully acknowledge the collaboration with 
D. Seipt, T. Nousch, T. Heinzl, 
and useful discussions with
A. Ilderton, K. Krajewska,  M. Marklund, C. M\"uller, S.~Rykovanov, 
R. Sch\"utzhold and G. Torgrimsson.
A. Ringwald is thanked for explanations w.r.t.\ LUXE.
The work is supported by R.~Sauerbrey and T.~E.~Cowan w.r.t.\ the study
of fundamental QED processes for HIBEF.

\end{document}